\documentclass[12pt]{article}
\usepackage[cp1251]{inputenc}

\usepackage{amsfonts,amssymb,amsthm,mathtools}
\usepackage{amsmath}
\usepackage{icomma}
\usepackage{latexsym}
\usepackage{graphicx}
\usepackage{subfigure}
\usepackage {bm}

\usepackage{geometry} 
\begin{document}
	
	\begin{center}
		\textbf{MATHEMATICAL PARADOXES OF DIRAC EQUATION REPRESENTATIONS}
	\end{center}
	
	\begin{center}
		
		{V.~P.~Neznamov\footnote{vpneznamov@mail.ru, vpneznamov@vniief.ru}}\\
		
		\hfil
		{\it \mbox{	Russian Federal Nuclear Center--All-Russian Research Institute of Experimental Physics},  Mira pr., 37, Sarov, 607188, Russia} \\
	\end{center}
	
\begin{abstract}
	\noindent
	\footnotesize{This paper examines the Foldy-Wouthuysen and Feynman-Gell-Mann 
		representations of the Dirac equation. The analysis is conducted for 
		electrons and positrons interacting with electromagnetic fields. Versions of 
		quantum electrodynamics are considered both within the scope of perturbation 
		theory and in the nonperturbativecase with strong electromagnetic fields. 
		Mathematical artifacts that contradicting the physical premises of the 
		theory are identified in the studied representations of the Dirac equation. 
		These mathematical paradoxes are resolved if the theory only employs 
		amplitude states (real and virtual) with positive energies.} \\
	
	
	
\end{abstract}


\section{Introduction}
The Dirac equation with a bispinor wave function is used in standard quantum 
electrodynamics (QED). The Dirac equation for an electron with mass $m$ and 
electric charge $e<0$, interacting with an electromagnetic field $A^{\mu 
}\left( {{\rm {\bf x}},t} \right)$, can be written in the form
\begin{equation}
\label{eq1}
p^{0}\psi_{D} \left( {{\rm {\bf x}},t} \right)=H_{D} \left( {{\rm {\bf 
x}},t} \right)\psi_{D} \left( {{\rm {\bf x}},t} \right)=\left( {{\rm {\bm 
\alpha }}\left( {{\rm {\bf p}}-e{\rm {\bf A}}\left( {{\rm {\bf x}},t} 
\right)} \right)+\beta m+eA^{0}\left( {{\rm {\bf x}},t} \right)} \right)\psi 
_{D} \left( {{\rm {\bf x}},t} \right).
\end{equation}
Here and below, the unit system $\hbar=c=1$ is used; $H_{D} \left( {{\rm 
{\bf x}},t} \right)$ is the Dirac Hamiltonian; $p^{\mu }=i (\partial 
/ \partial x_{\mu }) ,\,\,\,\mu =0,1,2,3$; $A^{\mu }\left( {{\rm {\bf x}},t} 
\right)$ are the electromagnetic potentials; and $\alpha^{i},\,\beta $ are 
the four-dimensional Dirac matrices. In the standard representation, the 
matrices $\alpha^{i},\,\,\,\beta ,\,\,\,\Sigma^{i},\gamma { }^{5},$ and $\gamma { 
}^{i}$ have the form
\begin{equation}
\label{eq2}
\begin{array}{l}
\alpha^{i}=\left( {\begin{array}{l}
 0\;\;\,\sigma^{i}{\kern 1pt} \\ 
 \sigma^{i}\;\;0 \\ 
 \end{array}} \right),\;\beta =\gamma^{0}=\left( {\begin{array}{l}
 I\;\;\,\,\,\,\,{\kern 1pt}0 \\ 
 0\;\,-I \\ 
 \end{array}} \right),\;\Sigma^{i}=\left( {\begin{array}{l}
 \sigma^{i}\;\;0\,{\kern 1pt} \\ 
 0\;\;\,\,\sigma^{i} \\ 
 \end{array}} \right), \\ [10pt]
\gamma^{5}=\left( {\begin{array}{l}
 0\;\;\,{\kern 1pt}I \\ 
 I\;\;\;0 \\ 
 \end{array}} \right),\;\gamma^{i}=\gamma^{0}\alpha^{i}.
\end{array}
\end{equation}
The bispinor $\psi_{D} \left( {{\rm {\bf x}},t} \right)$ can be written as
\begin{equation}
\label{eq3}
\psi_{D} ({\rm {\bf x}},t)=\left( {{\begin{array}{*{20}c}
 {\varphi ({\rm {\bf x}},t)} \hfill \\
 {\chi ({\rm {\bf x}},t)} \hfill \\
\end{array} }} \right).
\end{equation}
In a free case (without interaction), the Dirac equation has the following 
normalized solutions with positive and negative energies $\varepsilon $:
\begin{equation}
\label{eq4}
\begin{array}{l}
 \left( {\psi_{D} } \right)_{0}^{\left( + \right)} \left( {{\rm {\bf x}},t} 
\right)=\dfrac{1}{\left( {2\pi } \right)^{3/2}}\left( {1+\dfrac{{\rm {\bf 
p}}^{2}}{\left( {\left| E \right|+m} \right)^{2}}} \right)^{- 1/2}\left( {{\begin{array}{*{20}c}
 \,\,\,\,\,\,\,\,\,\,\,{U_{S} } \hfill \\ [5pt]
 {\dfrac{{\rm {\bm \sigma p}}}{\left| E \right|+m}\,\,U_{S} } \hfill \\
\end{array} }} \right)e^{-i\left| E \right|t+i{\rm {\bf 
px}}}, \\
\varepsilon =\left| E \right|>0, \\ 
 \left( {\psi_{D} } \right)_{0}^{\left( - \right)} \left( {{\rm {\bf x}},t} 
\right)=\dfrac{1}{\left( {2\pi } \right)^{3/2}}\left( {1+\dfrac{{\rm {\bf 
p}}^{2}}{\left( {\left| E \right|+m} \right)^{2}}} \right)^{-1/2} {\left( {{\begin{array}{*{20}c} 
 {\dfrac{{\rm {\bm \sigma p}}}{\left| E \right|+m}U_{S} } \hfill \\
 {\,\,\,\,\,\,\,\,\,\,\,\,U_{S} } \hfill \\ [5pt]
\end{array} }} \right)e^{i\left| E \right|t-i{\rm {\bf px}}}}, \\
\varepsilon =-\left| E \right|<0.
 \end{array}
\end{equation}
Here, $U_{S} $ are the normalized Pauli spinors $\left( {\mbox{for}\,\,S_{z} 
=1/2,\,\,U_{S} =\left( {{\begin{array}{*{20}c}
 1 \hfill \\
 0 \hfill \\
\end{array} }} \right)} \right., \\
\mbox{for}\,\,\,S_{z} =-1/2\left. {U_{S} =\left( {{\begin{array}{*{20}c}
 0 \hfill \\
 1 \hfill \\
\end{array} }} \right)} \right)$.

The Dirac equation also has solutions with positive and negative energies 
for stationary states in the presence of static electromagnetic fields.

On the one hand, the set of solutions with positive and negative energies 
provides mathematical completeness. On the other hand, the solutions with 
negative energies are not directly the solutions of the Dirac equation for 
antiparticles.The author of this equation, P.A.M. Dirac understood this well 
(see, e.g., \cite{bib1}). Two interpretations of the 
solutions with negative energy have become widely known.
\begin{enumerate}
	\item The physical vacuum of the Dirac equation is described using the concept 
	of fully occupied states with negative energies (the Dirac sea). Holes in 
	the Dirac sea are interpreted as antiparticles \cite{bib1}.
	\item In the Stueckelberg and Feynman positron theory \cite{bib2} - \cite{bib4}, positrons are 	electrons with negative energies that movein opposite directions in 
	spacetime.
\end{enumerate}

In standard QED, the fermion vacuum is nonvoid; virtual birth and 
annihilation of particles and antiparticles ar theoretically allowed in it.

There are versions of QED with a void fermion vacuum. These versions are 
based on using certain representations of the Dirac equation. They include 
the representation of Foldy and Wouthuysen (FW) \cite{bib5}, the representation of Feynman and Gell-Mann (FG) \cite{bib6}, and the representation with the Klein-Gordon (KG)-type fermion equations \cite{bib7}, \cite{bib8}.

For these representations, in the framework of perturbation theory, the 
formalisms (QED)$_{FW}$ \cite{bib9}, \cite{bib10}, (QED)$_{FG}$ \cite{bib11}, 
and (QED)$_{KG\, }$ \cite{bib12}, \cite{bib13} were developed, and some physical effects have been calculated. The final physical results fully agree with the respective results of standard QED with the Dirac equation.

Closed equations for fermions in the FW and FG representations were also 
been formulated for nonperturbative QED in strong electromagnetic 
fields \cite{bib14}, \cite{bib15}.

Using the a forementioned representations of the Dirac equation, it is 
sufficient to consider solutions with positive fermion energies when 
calculating physical effects. This applies to both real and intermediate 
virtual fermion states. In these cases, two separate equations are needed 
for fermions and antifermions. These equations differ by the sign of 
electric charge.

The use of these representations of the Dirac equatin to calculate QED 
effects revealed some contradictions between the physical premises of the 
theory and its mathematical results. All of these contradictions arise from 
the use of states with negativeenergies of fermion in calculations.

This paper analyzes the cause of these mathematical paradoxes.

It is organized as follows. Section 2 analyzes using (QED)$_{FW}$ in the 
applicability domain of perturbation theory. Section 3 presents an analysis 
for the representations of nonperturbative QED with strong electromagnetic 
fields. We consider the standard QED and the representations of 
Foldy-Wouthuysen and Feynman-Gell-Mann. We discuss the conclusions from our 
analysis in Section 4.
\section{Perturbation theory in Foldy-Wouthuysen representation}
\label{sec:mylabel1}
Two condition should hold in the Foldy-Wouthuysen representation (see, 
e.g., \cite{bib16}):
\begin{enumerate}
	\item The Hamiltonian or energy operators are diagonal with respect to the 
	upper and lower spinors of the wave function $\psi_{FW} ({\rm {\bf x}})$, 
	i.e., these operators do not mix the upper and lower components of $\psi 
	_{FW} ({\rm {\bf x}})$.
	\item The condition of wave function reduction holds under the Foldy-Wouthuysen 	transformation. When the Dirac Hamiltonian does not depend on time (the case of external static fields), the condition of reduction can be written as\footnote{Wave functions ar enormalized by unit probability in a box with volume $V$ below. For brevity, multipliers $1 / {\sqrt V }$ are absent from our expressions.}:
\end{enumerate}
\begin{equation}
\label{eq5}
\begin{array}{l}
\psi_{D}^{\left( + \right)} ({\rm {\bf x}},t)=e^{-i\varepsilon t}A_{\left( 
+ \right)} \left( {{\begin{array}{*{20}c}
 {\varphi ({\rm {\bf x}})} \hfill \\
 {\chi ({\rm {\bf x}})} \hfill \\
\end{array} }} \right) \to \\ [10pt]
\to \psi_{FW}^{\left( + \right)} ({\rm {\bf 
x}},t)=U_{FW}^{\left( + \right)} ({\rm {\bf x}})\psi_{D}^{\left( + \right)} 
({\rm {\bf x}},t)=e^{-i\varepsilon t}\left( {{\begin{array}{*{20}c}
 {\varphi ({\rm {\bf x}})} \hfill \\
 \,\,\,\,0 \hfill \\
\end{array} }} \right),
\end{array}
\end{equation}
where $\varepsilon >0$, and
\begin{equation}
\label{eq6}
\begin{array}{l}
\psi_{D}^{\left( - \right)} ({\rm {\bf x}},t)=e^{-i\varepsilon t}A_{\left( 
- \right)} \left( {{\begin{array}{*{20}c}
 {\varphi ({\rm {\bf x}})} \hfill \\
 {\chi ({\rm {\bf x}})} \hfill \\
\end{array} }} \right) \to \\ [10pt]
\to \psi_{FW}^{\left( - \right)} ({\rm {\bf 
x}},t)=U_{FW}^{\left( - \right)} ({\rm {\bf x}})\psi_{D}^{\left( - \right)} 
({\rm {\bf x}},t)=e^{-i\varepsilon t}\left( {{\begin{array}{*{20}c}
 \,\,\,0 \hfill \\
 {\chi ({\rm {\bf x}})} \hfill \\
\end{array} }} \right),
\end{array}
\end{equation}
where $\varepsilon <0$.

In (\ref{eq5}) and (\ref{eq6}), $A_{\left( + \right)} ,\,$ and $A_{\left( - \right)} $ are the normalizing operators, and $U_{FW}^{\left( + \right)} $ and $U_{FW}^{\left( -\right)} $ are the FW transformation operators. The operators $A_{\left( + \right)} ,\,\,A_{\left( - \right)} ,$ and $U_{FW}^{\left( + \right)} 
,\,\,\,U_{FW}^{\left( - \right)} $ are not necessarily the same for positive 
and negative energies.

In the FW representation, the Dirac equation for an electron interacting 
with an electromagnetic field $A^{\mu }\left( {{\rm {\bf x}},t} \right)$ can 
be obtained as a series in powers of the electromagnetic coupling constant 
by applying a series of unitary transformation of equation (\ref{eq1}) 
(see \cite{bib9}):
\begin{equation}
\label{eq7}
U_{FW} =\left( {1+e\delta_{1} +e^{2}\delta_{2} +e^{3}\delta_{3} +...} 
\right)U_{FW}^{0} .
\end{equation}
Here, $U_{FW}^{+} =U_{FW}^{-1} $.

As a result, we obtain the equation
\begin{equation}
\label{eq8}
\begin{array}{l}
 \varepsilon \psi_{FW} =H_{FW} \psi_{FW} =\left( {\beta E_{p} +eK_{1}^{FW} 
\left( {+m,A^{\mu }} \right)+} \right. \\ [10pt]
 \left. {+e^{2}K_{2}^{FW} \left( {+m,A^{\mu },A^{\nu }} 
\right)+e^{3}K_{3}^{FW} \left( {+m,A^{\mu },A^{\nu },A^{\gamma }} 
\right)+...} \right)\psi_{FW} . \\ 
 \end{array}
\end{equation}
Here, $E_{p} =\sqrt {m^{2}+{\rm {\bf p}}^{2}} $. The notation $+m$ in 
$K_{n}^{FW} $ indicates that the positive sign before $\beta m$ is taken in 
equation (\ref{eq1}). Equation (\ref{eq8}) does not contain terms with a negative sign before the mass $m$. This follows from the structure of the expressions for 
$K_{1}^{FW} ,\,\,K_{2}^{FW} \,...$ (see (39) -- (41), (\ref{eq20}), (\ref{eq21}) in \cite{bib9}).

In equation (\ref{eq8}),
\begin{equation}
\label{eq9}
\psi_{FW} =U_{FW} \psi_{D} .
\end{equation}
In the free case,
\begin{equation}
\label{eq10}
\varepsilon \left( {\psi_{FW} } \right)_{0} =\beta E_{p} \left( {\psi_{FW} 
} \right)_{0} ,
\end{equation}
where, for the positive energy $\varepsilon =\left| E \right|>0$,
\begin{equation}
\label{eq11}
\left( {\psi_{FW} } \right)_{0}^{\left( + \right)} \left( {{\rm {\bf x}},t} 
\right)=U_{FW}^{0} \left( {\psi_{D} } \right)_{0}^{\left( + \right)} 
=\frac{1}{\left( {2\pi } \right)^{3 \mathord{\left/ {\vphantom {3 2}} 
\right. \kern-\nulldelimiterspace} 2}}\left( {{\begin{array}{*{20}c}
 {U_{S} } \hfill \\
 \,\,\,0 \hfill \\
\end{array} }} \right)e^{-i\left| E \right|t+i{\rm {\bf px}}},
\end{equation}
and for the negative energy $\varepsilon =-\left| E \right|<0$,
\begin{equation}
\label{eq12}
\left( {\psi_{FW} } \right)_{0}^{\left( - \right)} \left( {{\rm {\bf x}},t} 
\right)=U_{FW}^{0} \left( {\psi_{D} } \right)_{0}^{\left( - \right)} 
=\frac{1}{\left( {2\pi } \right)^{3 \mathord{\left/ {\vphantom {3 2}} 
\right. \kern-\nulldelimiterspace} 2}}\left( {{\begin{array}{*{20}c}
 \,\,\,0 \hfill \\
 {U_{S} } \hfill \\
\end{array} }} \right)e^{i\left| E \right|t-i{\rm {\bf px}}}.
\end{equation}
In the FW representation, equation (\ref{eq8}) has noncovariant form, and the 
Hamiltonian $H_{FW} $ is nonlocal. Using standard methods of second 
quantization in quantum field theory is difficult in this case. However, the 
S-matrix approach and Feynman method of the propagator function \cite{bib2} - 
\cite{bib4}, \cite{bib17} can be used 
instead.In this method, QED processes are described by integral equations. 

The equation (\ref{eq8}) can be written for the four-dimensional $x,y$ in the form
\begin{equation}
\label{eq13}
\psi_{FW} \left( x \right)=\left( {\psi_{FW} } \right)_{0}^{\left( \pm 
\right)} \left( x \right)+\int {d^{4}y} \,S_{FW} \left( {x-y} 
\right)K^{FW} \left( y \right)\psi_{FW} \left( y \right),
\end{equation}
where $K^{FW} \left( y \right)=\sum\limits_{n=1}^\infty {e^{n}} K_{n}^{FW} 
\left( y \right)$ is the interaction Hamiltonian in equation (\ref{eq8}) and $S_{FW} 
\left( {x-y} \right)$ is the Feynman propagator in the Foldy-Wouthuysen 
representation,
\begin{equation}
\label{eq14}
S_{FW} \left( {x-y} \right)=\frac{1}{\left( {2\pi } \right)^{4}}\int 
{d^{4}p} \,e^{-ip\left( {x-y} \right)}\frac{p^{0}+\beta 
E}{p^{2}-m^{2}+i\varepsilon }.
\end{equation}
The elementsof the $S$-matrix can be written as
\begin{equation}
\label{eq15}
S_{fi} =\delta_{fi} -i\varepsilon_{f} \int {d^{4}y} \,\left[ {\left( 
{\bar{{\psi }}_{FW} } \right)_{0}^{\left( \pm \right)} \left( y \right)} 
\right]_{f} K^{FW} \left( y \right)\left[ {\psi_{FW} \left( y \right)} 
\right]_{i} .
\end{equation}
Here, the bar over the function $\psi_{FW} $ denotes the Hermitian 
conjugate, $\varepsilon_{f} =+1$ for the solution $\,\left[ {\left( 
{\bar{{\psi }}_{FW} } \right)_{0}^{\left( + \right)} \left( y \right)} 
\right]_{f} $ and $\varepsilon_{f} =-1$ for the solution$\,\left[ {\left( 
{\bar{{\psi }}_{FW} } \right)_{0}^{\left( - \right)} \left( y \right)} 
\right]_{f} $.

For the standard QED, equations (\ref{eq13}) and (\ref{eq15}) were derived in the works of R.~Feynman \cite{bib3}, \cite{bib4} and also, 
for example,in monograph \cite{bib17}.

Additionally, we mention several important points.
\begin{enumerate}
	\item The Hamiltonians $H_{FW} $ in (\ref{eq8}) and $K^{FW} \left( y \right)$ in (\ref{eq13}) are diagonal with respect to the mixing of the upper and lower components of the bispinor $\psi_{FW} $. Each of the equations (\ref{eq8}) and (\ref{eq13}) includes two independent equations with the spinor wave functions $\sim U_{S} $. One equation contains the states with positive energies, and the other one contains the states with negative energies. The elements of the S-matrix in (\ref{eq15}) can be calculated by using only the states with positive energy. In this case, the states with negative energy are not used in calculations of physical processes in the QED: they are only needed for mathematical completeness in the expansion of operators and wave functions.
	\item In standard QED with the Dirac equation, positrons are electrons with 
	negative energies moving in opposite directions in spacetime. In the 
	Foldy-Wouthuysen representation, the situation changes. If in (\ref{eq15}) we use 
	$\left[ {\left( {\bar{{\psi }}_{FW} } \right)_{0}^{\left( + \right)} } 
	\right]_{f} $ on the left-hand side and $-\left[ {\left( {\bar{{\psi}}_{FW} } 
		\right)_{0}^{\left( - \right)} } \right]_{i} $ on the right-hand side, then due to the structure of the bispinors (\ref{eq11}), (\ref{eq12}), the respective elements of the S-matrix will be zero in all orders of perturbation theory. Taking (\ref{eq11}) and (\ref{eq12}) into account, we can write
\end{enumerate}
\[
\left\langle {\frac{1}{\left( {2\pi } \right)^{3/2}}e^{i\left| {E_{f} } 
\right|t\,-\,i{\rm {\bf p}}_{{\rm {f}}} {\rm {\bf x}}}\left( 
{\,{\begin{array}{*{20}c}
 {\bar{{U}}_{S} } \hfill & 0 \hfill \\
\end{array} }\,} \right)\,\,\left| {M_{FW} } \right|\,\,\left( 
{{\begin{array}{*{20}c}
\,\, 0 \hfill \\
 {U_{S} } \hfill \\
\end{array} }} \right)\frac{1}{\left( {2\pi } \right)^{3/2}}e^{i\left| {E_{i} } 
\right|t\,-\,i{\rm {\bf p}}_{{\rm {i}}} {\rm {\bf x}}}} \right\rangle 
=0.
\]
Here, by definition, $M_{FW} $ is a diagonal operator.

A similar result is obtained when in (\ref{eq15}) we use $\left[ {\left( {\bar{{\psi }}_{FW} } \right)_{0}^{\left( - \right)} } \right]_{f} $ on the left-hand side and $-\left[ {\left( {\bar{{\psi}}_{FW} } \right)_{0}^{\left( + \right)} } 
\right]_{i} $ on the right-hand side.

Thus, positrons in the FW representation cannot be described by electron 
states with negative energy. Positrons in this representation should be 
described by positiveenergy states of a special equation for positrons.

We have therefore obtained the first paradox. \textit{By performing a unitary transformation of the Dirac equation in the FW representation, we lost the interaction between the states with positive and negative energies.}
\section{Nonperturbative quantum electrodynamics and representations of Dirac equation}
\label{sec:mylabel2}
Consider the case of electrostatic fields of hydrogen-like ions.
\subsection{Standard quantum electrodynamics in fields of hydrogen-like ions with large charge number Z}
Figure 1 shows the lower energy levels of a hydrogen-like ion as a function 
of the charge number Z for the standard QED with a fluctuating fermionic 
vacuum. This figure is adapted from monograph \cite{bib18}.

\begin{figure}[h!]
	\centerline{\includegraphics[width=0.6\linewidth]{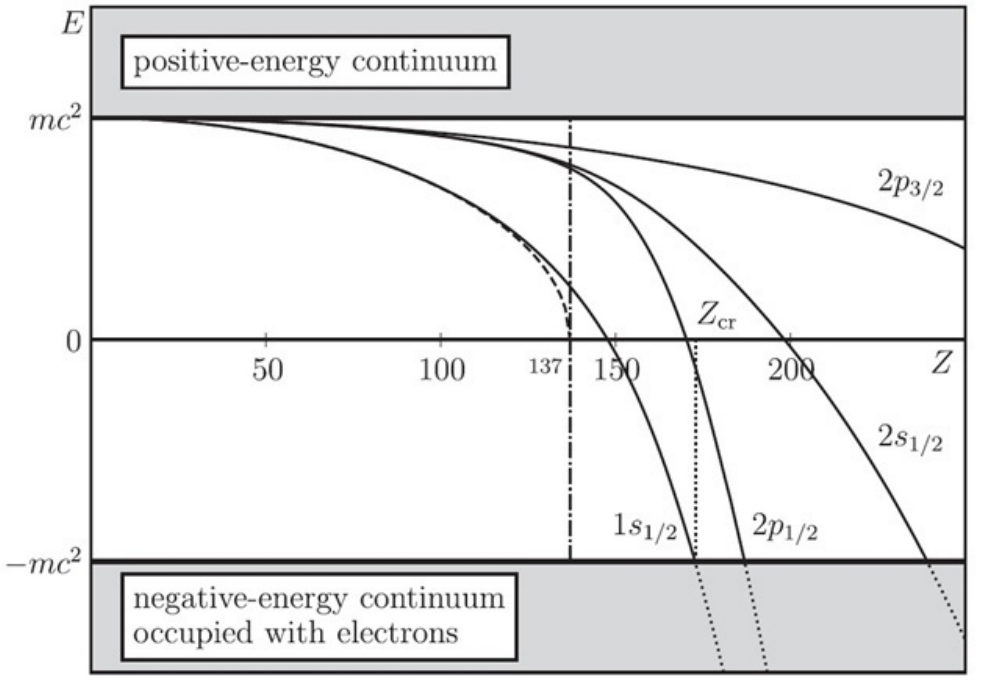}}
	\caption{Lower energy levels of hydrogen-like ion as a function of nuclear 
		charge number $Z$.}
	\label{ris:Fig.1}
\end{figure}

Consider the level $1s_{1/2} $. For the Coulomb field of apoint nucleus 
charge $+Z|e|$, the level $1s_{1/2} $ disappears on reaching $Z=137$. If finite 
dimensions of atomic nuclei are considered \cite{bib19} - \cite{bib22}, the energy of the $1s_{1/2}$ becomes negative for $Z>146$. For $Z_{cr} \approx 171$, the level $1s_{1/2} $ becomes immersed in a negative-energy continuum. Similarly, the energy of the level $2p_{1/2}$ becomes negative for $Z>168$; for $Z_{cr} \approx 184$, level $2p_{1/2}$ becomes immersed in the negative- energy continuum. According to theoretical predictions of standard QED, when a level becomes immersed in the negative-energy continuum, the neutral vacuum decays, emitting two 
electron-positron pairs \cite{bib20}, \cite{bib21}.

\subsection{Representations of Feynman-Gell-Mann and Foldy-Wouthuysen}
\label{sec:mylabel3}
We write the Dirac equation in an external electromagnetic field in 
covariant form:
\begin{equation}
\label{eq16}
\left( {\gamma^{0}\left( {p^{0}-eA^{0}} \right)-{\rm {\bf \gamma }}\left( 
{{\rm {\bf p}}-e{\rm {\bf A}}} \right)-m} \right)\psi_{D} ({\rm {\bf 
x}},t)=0.
\end{equation}
We multiply the left-hand side of (\ref{eq16}) by the operator with the changed sign of the fermion mass,
\begin{equation}
\label{eq17}
\begin{array}{l}
\left( {\gamma^{0}\left( {p^{0}-eA^{0}} \right)-{\rm {\bf \gamma }}\left( 
{{\rm {\bf p}}-e{\rm {\bf A}}} \right)+m} \right)\times \\ [5pt]
\times \left( {\gamma^{0}\left( 
{p^{0}-eA^{0}} \right)-{\rm {\bf \gamma }}\left( {{\rm {\bf p}}-e{\rm {\bf 
A}}} \right)-m} \right)\psi_{D} ({\rm {\bf x}},t)=0.
\end{array}
\end{equation}
The result is a second order equation whose solutions are degenerate with 
respect to the sign before the mass $m$:
\begin{equation}
\label{eq18}
\left[ {\left( {p^{0}-eA^{0}} \right)^{2}-\left( {{\rm {\bf p}}-e{\rm {\bf 
A}}} \right)^{2}-m^{2}+e{\rm {\bf \Sigma H}}-ie{\rm {\bm \alpha E}}} 
\right]\psi_{D} ({\rm {\bf x}},t)=0.
\end{equation}
Here, ${\rm {\bf H}}=rot{\rm {\bf A}},\,\,{\rm {\bf E}}=-(\partial \rm 
{\bf A}/ \partial t)-\vec{{\nabla }}A^{0}$ are the magnetic and electric 
fields, respectively.

We limit ourselves to the case of static electromagnetic fields 
when $p^{0}\psi =\varepsilon \psi$.

To transition to the FG representation, one needs to use the Dirac matrices 
in the chiral representation in equation (\ref{eq18}). This is achieved by 
the unitary transformation
\begin{equation}
\label{eq19}
S=\,S^{-1}=\frac{1}{\sqrt 2 }\left( {{\begin{array}{*{20}c}
 I \hfill & \,\,\,I \hfill \\
 I \hfill & {-I} \hfill \\
\end{array} }} \right),
\end{equation}
\begin{equation}
\label{eq20}
\psi_{FG} \left( {{\rm {\bf x}},t} \right)=S\psi \left( {{\rm {\bf x}},t} 
\right)=\left( {{\begin{array}{*{20}c}
 {\varphi_{FG} \left( {{\rm {\bf x}}} \right)} \hfill \\
 {\chi_{FG} \left( {{\rm {\bf x}}} \right)} \hfill \\
\end{array} }} \right)e^{-i\varepsilon t},
\end{equation}
\begin{equation}
\label{eq21}
\alpha_{c}^{i} =S\alpha^{i} S^{-1}=\left( {{\begin{array}{*{20}c}
 {\sigma^{i}} \hfill & \,\,\,\,0 \hfill \\
 0 \hfill & {-\sigma^{i}} \hfill \\
\end{array} }} \right),\;\,\,\Sigma_{c} =S\Sigma^{i} S^{-1}=\left( 
{\begin{array}{l}
 \sigma^{i}\;\;0\,{\kern 1pt} \\ 
 0\;\;\,\,\sigma^{i} \\ 
 \end{array}} \right).
\end{equation}
In the FG representation, there is no mixing of the upper and lower 
components of the bispino r$\psi_{FG} \left( {{\rm {\bf x}},t} \right)$ 
inequation. In the case of stationary states, equation (\ref{eq18}) is reduced to two separate equations for spinors $\varphi_{FG} \left( {{\rm {\bf x}}} 
\right)$ and $\chi_{FG} \left( {{\rm {\bf x}}} \right)$:
\begin{equation}
\label{eq22}
\left[ {\left( {\varepsilon -eA^{0}} \right)^{2}-\left( {{\rm {\bf p}}-e{\rm 
{\bf A}}} \right)^{2}-m^{2}+e{\rm {\bm \sigma \bf { H}}}-ie{\rm {\bm \sigma \bf {E}}}} 
\right]\varphi_{FG} ({\rm {\bf x}})=0,
\end{equation}
\begin{equation}
\label{eq23}
\left[ {\left( {\varepsilon -eA^{0}} \right)^{2}-\left( {{\rm {\bf p}}-e{\rm 
{\bf A}}} \right)^{2}-m^{2}+e{\rm {\bm \sigma \bf {H}}}+ie{\rm {\bm \sigma \bf { E}}}} 
\right]\chi_{FG} ({\rm {\bf x}})=0.
\end{equation}
Analogous equations were considered earlier by Feynman and Gell-Mann \cite{bib6}.

It is noteworthy that equations (\ref{eq22}) and (\ref{eq23}) are connected with equations in the Foldy-Wouthuysen representation \cite{bib14}, \cite{bib15}. Equation (\ref{eq22}) in the FW representation is obtained for positive energies $\varepsilon =\left| E \right|>0$. In this case, $\varphi_{FG} \left( {{\rm {\bf x}}} \right)=A_{\left( + \right)} 
\,\varphi_{c} \left( {{\rm {\bf x}}} \right)$, where $\varphi_{c} \left( 
{{\rm {\bf x}}} \right)$ is the upper spinor in the Foldy-Wouthuysen 
representation, $ {\psi_{FW}^{\left( + \right)} ({\rm {\bf x}})=\left( 
{{\begin{array}{*{20}c}
 {\varphi_{c} ({\rm {\bf x}})} \hfill \\
 \,\,\,\,\,0 \hfill \\
\end{array} }} \right)}$. Equation (\ref{eq23}) in the FW representation is obtained for negative energies $\varepsilon =-\left| E \right|<0$. In this 
case, $\chi_{FG} \left( {{\rm {\bf x}}} \right)=A_{\left( + \right)} \,\chi 
_{c} \left( {{\rm {\bf x}}} \right)$, where $\chi_{c} \left( {{\rm {\bf 
x}}} \right)$ is the lower spinor in the FW representation, ${\psi 
_{FW}^{\left( - \right)} ({\rm {\bf x}})=\left( {{\begin{array}{*{20}c}
 \,\,\,\,\,0 \hfill \\
 {\chi_{c} ({\rm {\bf x}})} \hfill \\
\end{array} }} \right)}$.

We write, in agreement with \cite{bib14}, certain equations 
for electrons and positrons in the FG and FW representations.

\begin{enumerate}
	\item Equation for electrons with positive energies $\left( {\varepsilon =\left| 
		E \right|>0,\,\,\,e=-\left| e \right|<0} \right)$
	\begin{equation}
		\label{eq24}
		\left( {\left( {\left| E \right|+\left| e \right|A^{0}} \right)^{2}-{\rm 
				{\bf p}}^{2}-m^{2}-i\left| e \right|{\rm {\bm \sigma }}\,\vec{{\nabla 
			}}A^{0}} \right)\varphi_{FG}^{e} =0.
	\end{equation}
	\item Equation for electrons with negative energies $\left( {\varepsilon 
		=-\left| E \right|<0,\,\,\,e=-\left| e \right|<0} \right)$
	\begin{equation}
		\label{eq25}
		\left( {\left( {\left| E \right|-\left| e \right|A^{0}} \right)^{2}-{\rm 
				{\bf p}}^{2}-m^{2}+i\left| e \right|{\rm {\bm \sigma }}\,\vec{{\nabla 
			}}A^{0}} \right)\chi_{FG}^{e} =0.
	\end{equation}
	\item Equation for positrons with positive energies $\left( {\varepsilon =\left| 
		E \right|>0,\,\,\,e=\left| e \right|>0} \right)$
	\begin{equation}
		\label{eq26}
		\left( {\left( {\left| E \right|-\left| e \right|A^{0}} \right)^{2}-{\rm 
				{\bf p}}^{2}-m^{2}+i\left| e \right|{\rm {\bm \sigma }}\,\vec{{\nabla 
			}}A^{0}} \right)\varphi_{FG}^{p} =0.
	\end{equation}
\end{enumerate}

In equation (\ref{eq24}), the spinor $\varphi_{FG}^{e} \left( {{\rm {\bf x}}} 
\right)$ in the Feynman-Gell-Mann representation is proportional to the 
upper spinor $\varphi_{c}^{e} \left( {{\rm {\bf x}}} \right)$ in the 
Foldy-Wouthuysen representation:
\begin{equation}
\label{eq27}
\varphi_{FG}^{e} \left( {{\rm {\bf x}}} \right)=A_{\left( + \right)}^{e} 
\varphi_{c}^{e} \left( {{\rm {\bf x}}} \right);\,\,\,\left( {\psi 
_{FW}^{\left( + \right)} \left( {{\rm {\bf x}}} \right)} \right)^{e}=\left( 
{{\begin{array}{*{20}c}
 {\varphi_{c}^{e} \left( {{\rm {\bf x}}} \right)} \hfill \\
 \,\,\,\,\,0 \hfill \\
\end{array} }} \right).
\end{equation}
In equation (\ref{eq25}), the spinor $\chi_{FG}^{e} \left( {{\rm {\bf x}}} \right)$ in the Feynman-Gell-Mann representation is proportional to the lower spinor $\chi_{c}^{e} \left( {{\rm {\bf x}}} \right)$ in the 
Foldy-Wouthuysen representation:
\begin{equation}
\label{eq28}
\chi_{FG}^{e} \left( {{\rm {\bf x}}} \right)=A_{\left( - \right)}^{e} \chi 
_{c}^{e} \left( {{\rm {\bf x}}} \right);\,\,\,\left( {\psi_{FW}^{\left( - 
\right)} \left( {{\rm {\bf x}}} \right)} \right)^{e}=\left( 
{{\begin{array}{*{20}c}
 \,\,\,\,\,0 \hfill \\
 {\chi_{c}^{e} \left( {{\rm {\bf x}}} \right)} \hfill \\
\end{array} }} \right).
\end{equation}
In equation (\ref{eq26}), the spinor $\varphi_{FG}^{p} \left( {{\rm {\bf x}}} 
\right)$ in the Feynman-Gell-Mann representation is proportional to the 
upper spinor $\varphi_{c}^{p} \left( {{\rm {\bf x}}} \right)$ in the 
Foldy-Wouthuysen representation:
\begin{equation}
\label{eq29}
\varphi_{FG}^{p} \left( {{\rm {\bf x}}} \right)=A_{\left( + \right)}^{p} 
\varphi_{c}^{p} \left( {{\rm {\bf x}}} \right);\,\,\,\left( {\psi 
_{FW}^{\left( + \right)} \left( {{\rm {\bf x}}} \right)} \right)^{p}=\left( 
{{\begin{array}{*{20}c}
 {\varphi_{c}^{p} \left( {{\rm {\bf x}}} \right)} \hfill \\
 \,\,\,\,\,0 \hfill \\
\end{array} }} \right).
\end{equation}
Here, 
\begin{equation*}
A_{\left( + \right)}^{e} =\left( {1+\frac{m^{2}}{\left( {\left| E 
			\right|+{\rm {\bm \sigma \bf {p}}}+\left| e \right|A^{0}} \right)^{2}}} 
\right)^{-1/2},
\end{equation*}
\begin{equation*}
	A_{\left( - \right)}^{e} =\left( 
	{1+\frac{m^{2}}{\left( {\left| E \right|+{\rm {\bm \sigma \bf {p}}}-\left| e 
				\right|A^{0}} \right)^{2}}} \right)^{-1/ 2},
\end{equation*}
\begin{equation*}
	A_{\left( + \right)}^{p} =\left( 
	{1+\frac{m^{2}}{\left( {\left| E \right|+{\rm {\bm \sigma \bf {p}}}-\left| e 
				\right|A^{0}} \right)^{2}}} \right)^{-1/ 2}.
\end{equation*}

As a result, we see that equation (\ref{eq26}) for positrons with positive energies $\varepsilon >0$ coincides with equation (\ref{eq25}) for electrons with negative energies $\varepsilon <0$. This conclusion is confirmed by separating the 
variables. The system of equations for the radial functions, derived from 
equations (\ref{eq25}) and (\ref{eq26}), are fully consistent with each other.

In our case, the positrons are in the repulsive Coulomb field of ionized 
nuclei. For them, an upper continuum exists with a continuous energy 
spectrum $\varepsilon >m$. However, stationary bound states with 
$\varepsilon <m$ are absent in this case.

The equality of equations (\ref{eq25}) and (\ref{eq26}) assumes that electrons with negative energies and positrons with positive energies have an equivalent (up to the sign of energy) continuous energy spectrum with the same stationary wave 
functions. However, the spectrum of equation (\ref{eq25}) contains the negative 
energy level $1s_{1/2}$ in the interval $Z_{\Sigma } =147-170$ 
and the negative energy level $2p_{1/2}$ in the interval $Z_{\Sigma } =169-183$ (see Fig. 1).

Simple physical considerations prohibit such bound states for equation (\ref{eq26}). 
Therefore, the existence of bound states with negative energies is a 
mathematical artifact.

Since equations (\ref{eq24}) - (\ref{eq26}) are obtained through unitary transformations of the Dirac equation, the conclusion that there is no physical (and not mathematical) contribution of bound states with negative energies in effects calculated in QED is also valid for the original Dirac equation.

Note that equation (\ref{eq26}) with a changed sign by $A^{0}\left( {{\rm {\bf x}}} \right)$ (the motion of a positron in an attractive Coulomb field) coincides 
with equation (\ref{eq24}) for electrons. As expected, discrete and continuous 
energy spectra of electrons and positrons moving in an attractive Coulomb 
field coincide with each other.

\subsection{Dirac equation with nonrelativistic Hamiltonian in Foldy-Wouthuysen representation}
\label{sec:mylabel4}
The conclusions in Section 3.2 are confirmed by analyzing the 
nonrelativistic Hamiltonian in the FW representation, which was already 
obtained in initial work on the Foldy-Wouthuysen transformation (see \cite{bib5}).

Consider nonrelativistic motion of electrons and positrons in an external 
electrostatic field $eA^{0}\left( {{\rm {\bf x}}} \right)$. According to \cite{bib5}, the nonrelativistic Hamiltonian $H_{FW} $ takes 
the form
\begin{equation}
\label{eq30}
H_{FW} =\beta \left( {m+\frac{{\rm {\bf p}}^{2}}{2m}} 
\right)+eA^{0}-\frac{ie}{8m^{2}}{\rm {\bm \sigma }}\,\mbox{rot}\,{\rm {\bf 
E}}-\frac{e}{4m^{2}}{\rm {\bm \sigma }}\,\left( {{\rm {\bf E}}\times {\rm 
{\bf p}}} \right)-\frac{e}{8m^{2}}\,\mbox{div}\,{\rm {\bf E}}.
\end{equation}
Here, ${\rm {\bf E=}}-\vec{{\nabla }}A^{0}$ is the electric field.

The Dirac equation with the Hamiltonian $H_{FW} $ is written in the form
\begin{equation}
\label{eq31}
\begin{array}{l}
\left[ {\left( {\varepsilon -eA^{0}} \right)-\beta \left( {m+\dfrac{{\rm {\bf 
p}}^{2}}{2m}} \right)+\dfrac{ie}{8m^{2}}{\rm {\bm \sigma }}\,\mbox{rot}\,{\rm 
{\bf E}}+\dfrac{e}{4m^{2}}{\rm {\bm \sigma }}\,\left( {{\rm {\bf E}}\times 
{\rm {\bf p}}} \right)+\dfrac{e}{8m^{2}}\,\mbox{div}\,{\rm {\bf E}}} 
\right] \times \\ [10pt]
\times \psi_{FW} \left( {{\rm {\bf x}},t} \right)=0.
\end{array}
\end{equation}
Here, for $\varepsilon >0$, 

$\psi_{FW}^{\left( + \right)} \left( {{\rm {\bf 
x}},t} \right)=\left( {{\begin{array}{*{20}c}
 {\varphi \left( {{\rm {\bf x}}} \right)} \hfill \\
\,\,\,\,0 \hfill \\
\end{array} }} \right)e^{-i\varepsilon t}$, and, in this case $\beta \psi 
_{FW}^{\left( + \right)} \left( {{\rm {\bf x}},t} \right)=\psi_{FW}^{\left( 
+ \right)} \left( {{\rm {\bf x}},t} \right)$;

for $\varepsilon <0$,

$\psi_{FW}^{\left( - \right)} \left( {{\rm {\bf x}},t} 
\right)=\left( {{\begin{array}{*{20}c}
 \,\,\,\,0 \hfill \\
 {\chi \left( {{\rm {\bf x}}} \right)} \hfill \\
\end{array} }} \right)e^{-i\varepsilon t}$, and, in this case $\beta \psi 
_{FW}^{\left( - \right)} \left( {{\rm {\bf x}},t} \right)=-\psi 
_{FW}^{\left( - \right)} \left( {{\rm {\bf x}},t} \right)$.

Just as in Section 3.2, consider equations (\ref{eq31}) for electrons and positrons.
\begin{enumerate}
	\item The equation for electrons with positive energies $\left( {\varepsilon 
		=\left| E \right|>0,\,\,\,e=-\left| e \right|<0} \right)$:
	\begin{equation}
		\label{eq32}
		\begin{array}{l}
			\left[ {\left( {\left| E \right|+\left| e \right|A^{0}} \right)-\left( 
			{m^{2}+\dfrac{{\rm {\bf p}}^{2}}{2m}} \right)-\dfrac{i\left| e 
				\right|}{8m^{2}}{\rm {\bm \sigma }}\mbox{rot}{\rm {\bf E}}-\dfrac{\left| 
				{{e}} \right|}{4m^{2}}{\rm {\bm \sigma }}\left( {{\rm {\bf 
						E}}\times {\rm {\bf p}}} \right)-\dfrac{\left| {{e}} 
				\right|}{8m^{2}}\mbox{div}\,{\rm {\bf E}}} \right] \times \\ [5pt]
			\times \varphi^{e}\left( {{\rm 
				{\bf x}}} \right)=0.
		\end{array}
	\end{equation}
	\item The equation for electrons with negative energies $\left( {\varepsilon 
		=-\left| E \right|<0,\,\,\,e=-\left| e \right|<0} \right)$:
	\begin{equation}
		\label{eq33}
		\begin{array}{l}
			\left[ {\left( {\left| E \right|-\left| e \right|A^{0}} \right)-\left( 
			{m^{2}+\dfrac{{\rm {\bf p}}^{2}}{2m}} \right)+\dfrac{i\left| e 
				\right|}{8m^{2}}{\rm {\bm \sigma }}\,\mbox{rot}\,{\rm {\bf E}}+\dfrac{\left| 
				{{e}} \right|}{4m^{2}}{\rm {\bm \sigma }}\,\left( {{\rm {\bf 
						E}}\times {\rm {\bf p}}} \right)+\dfrac{\left| {{e}} 
				\right|}{8m^{2}}\mbox{div}\,{\rm {\bf E}}} \right] \times \\ [5pt]
			\times \chi^{e}\left( {{\rm {\bf x}}} \right)=0.
				\end{array}
	\end{equation}
	\item The equation for positrons with positive energies $\left( {\varepsilon 
		=\left| E \right|>0,\,\,\,e=\left| e \right|>0} \right)$:
	\begin{equation}
		\label{eq34}
		\begin{array}{l}
		\left[ {\left( {\left| E \right|-\left| e \right|A^{0}} \right)-\left( 
			{m^{2}+\dfrac{{\rm {\bf p}}^{2}}{2m}} \right)+\dfrac{i\left| e 
				\right|}{8m^{2}}{\rm {\bm \sigma }}\,\mbox{rot}\,{\rm {\bf E}}+\dfrac{\left| 
				{{e}} \right|}{4m^{2}}{\rm {\bm \sigma }}\,\left( {{\rm {\bf 
						E}}\times {\rm {\bf p}}} \right)+\dfrac{\left| {{e}} 
				\right|}{8m^{2}}\mbox{div}\,{\rm {\bf E}}} \right] \times \\ [5pt] \times \varphi^{p}\left( {{\rm 
				{\bf x}}} \right)=0.
				\end{array}
	\end{equation}
\end{enumerate}

It can be seen that equations (\ref{eq33}) and (\ref{eq34}) coincide, i.e., just as in the preceding sections,the equations for positrons with $\varepsilon >0$ 
coincides with the equation for electrons with $\varepsilon <0$ in the 
Coulomb field of atomic nuclei.

However, in the nonrelativistic case with $\left| E \right|-m\ll 1$, the 
coincidence of equations (\ref{eq33}) and (\ref{eq34}) does not lead to a conflict with physical reality. Indeed, in this case, equation (\ref{eq33}) does not produce discrete levelswith negative energies (see Fig. 1).

Such levels arise mathematically in the domain of strong electrostatic 
fields with $Z>146$. This contradicts clear physical arguments about the 
absence of the discrete levels with negative energies.

\section{Conclusions}
\label{sec:mylabel5}
Certain paradoxes of the Dirac equation have been explained in the FW 
representation earlier.

\begin{enumerate}
	\item In the Dirac equation without interaction, the velocity of fermions is 
	${\rm {\bf v}}_{D} =c{\rm {\bm \alpha }}$ \cite{bib1}. In 
	the FW representation, the velocity of fermions takes the classical 
	form ${\rm {\bf v}}_{FW} ={c{\rm {\bf p}}} /E$ \cite{bib5}.
	\item In the FW representation, the ''Zitterbewegung'' of the fermion 
	coordinatesis absent \cite{bib5}. See also \cite{bib23}, \cite{bib24} on the problem of Zitterbewegung.
	\item As many authors have mentioned, a positive aspect of the FW representation 	is that the correspondence principle between quantum-mechanical operators and analogous classical quantities is explicitly observed in nonrelativistic quantum mechanics. As can be seen, for example, from pp. 1, 2, such a correspondence is often absent in the Dirac representation.
\end{enumerate}

These facts are related to the absence of virtual interactions between 
fermions with positive and negative energies in the FW representation.

The physical community has become accustomed to theparadoxes of the Dirac 
equation, because they do not affect the results of calculations of physical 
effects in QED.

Our analysis revealed two new paradoxes, whose resolution influences the 
physical effects of QED.

\textit{Paradox } No. 1. After applying a unitary transformation to the Dirac equation in an external electromagnetic field and transition to the FW representation, the elements of the S-matrix lose the interaction between the states with positive and negative energy. To restore the interaction, an additional 
equation for positrons with positive energies needs to be introduced into 
the theory.

\textit{Paradox} No. 2. Contrary to clear a clear physical picture, mathematical calculations of the energy spectrum for hydrogen-like ions with $Z_{\Sigma } =147-183$ produce levels with negative energies.\footnote{Discrete levels with 
negative energies appear in calculations that consider the finite size of 
nuclei \cite{bib19} - \cite{bib22}. In calculations with the Coulomb field of point nucleus charge $+Z|e|$, such states are absent.}

Both the paradoxes are related to negative-energy solutions of the Dirac 
equation.

All paradoxes disappearwhen only states with positiveenergies are used to 
calculate the physical effects of QED.The states with negative energy should 
only be taken into account to ensure completeness in the expansions of wave 
functions and operators. In this case, in addition to the equation for 
electrons with positive energies, an equation for positrons with positive 
energies is introduced.

This approach applies to (QED)$_{FW}$, (QED)$_{KG}$, and also for the 
standard QED with the Dirac equation (see \cite{bib25}). Within the domain of applicability of perturbation theory, the final physical results coincide with those of standard QED.

In nonperturbative QED, one should consider the spectrum shown in Fig. 2 
instead of spectrum of hydrogen-like ions with the charges $Z\left| e 
\right|$ (see Fig. 1).
\begin{figure}[h!]
	\begin{minipage}[h]{0.47\linewidth}
		\center{\includegraphics[width=1\linewidth]{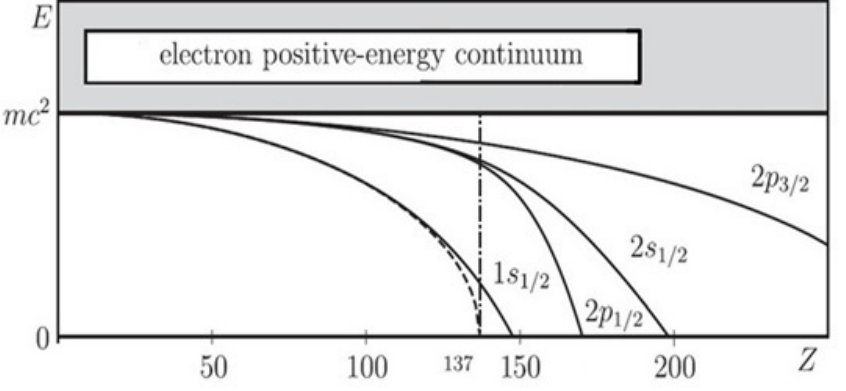}} a) \\
	\end{minipage}
	\hfill
	\begin{minipage}[h]{0.47\linewidth}
		\center{\includegraphics[width=1\linewidth]{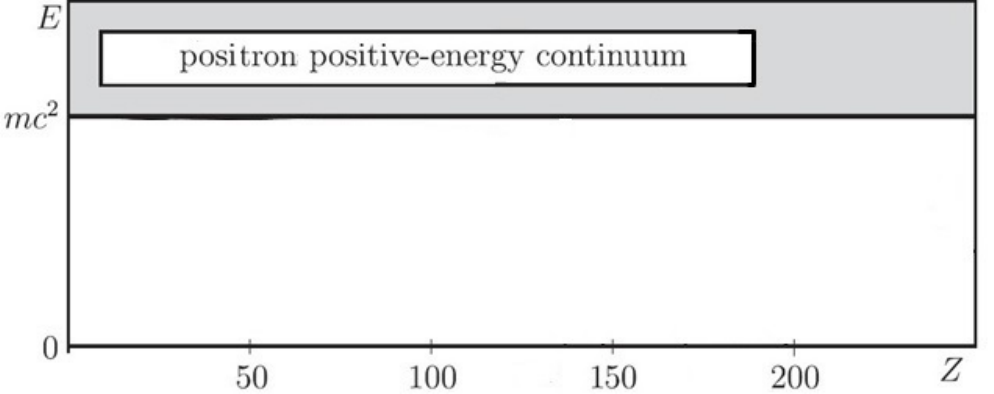}} \\b) 
	\end{minipage}
	\caption{Energy spectrum of a) equation for electrons (\ref{eq24}) and b) equation for positrons (\ref{eq25}).}
	\label{ris:Fig.2}
\end{figure}

In nonperturbative QED, the calculated spectrum shown in Fig. 2a (without 
discrete and continuous spectra with negative energies) can be confirmed 
experimentally. Reference \cite{bib26} suggests performing 
a series of experiments on heavy ion colliders to confirm this.

We note that P.~A.~M.~Dirac, dissatisfied with the presence of states with 
negative energies in his equation, already at the end of his life turned to 
searching for a relativistic wave equation with only positive energy 
solutions \cite{bib27}, \cite{bib28}. 
However, he did not succeed in carrying his ideas to a logical end. In this 
work, we continue the path to solving physical problems in quantum 
electrodynamics caused by the presence of fermion states with negative 
energies.

\section{Acknowledgements}
This work was carried out in the framework of the scientific program at the 
National Center for Physics and Mathematics, direction ''Particle physics 
and cosmology. Stage 2023-2025''.

The author would like to thank A.L.Novoselova for help with the preparation 
of this work.



\begin{thebibliography}{00}  
	
\bibitem{bib1} P.~A.~M.Dirac, {\it The Principles of Quantum Mechanics} (Oxford:The Univ. Press, 1930).

\bibitem{bib2} E.~C.~G.~Stuekelberg, {\it Helv. Phys. Acta.} {\bf 14} L32 (1941); {\it Helv. Phys. Acta.} {\bf14}, 588 (1941).

\bibitem{bib3} R.~P.~Feynman, {\it Phys. Rev.} {\bf 76}, 749 (1949).

\bibitem{bib4} R.~P.~Feynman, {\it Phys. Rev.} {\bf76}, 769 (1949).

\bibitem{bib5} L.~L.~Foldy, S.~A.~Wouthuysen, {\it Phys. Rev.} {\bf 78}, 29 (1950).

\bibitem{bib6} R.~P.~Feynman and M.~Gell-Mann, {\it Phys. Rev.} {\bf109}, 193 (1958).

\bibitem{bib7} Y.~B.~Zel'dovich and V.~S.~Popov, {\it Sov. Phys. Usp}. {\bf 14}, 673 (1972).

\bibitem{bib8} V.~P.~Neznamov, I.~I.~Safronov, {\it J. Exp. Theor. Phys.} {\bf128}, 672 (2019), arxiv: 1907.03579.

\bibitem{bib9} V.~P.~Neznamov, {\it Phys.Part. Nucl}. {\bf37}, 86 (2006);arxiv: hep-th/0411050.

\bibitem{bib10} V.~P.~Neznamov, {\it Phys.Part. Nucl. } {\bf43}, 36 (2012);arxiv: 1107.0693.

\bibitem{bib11} L.~M.~Brown, {\it Phys. Rev.} {\bf111}, 957 (1958).

\bibitem{bib12} V.~P.~Neznamov, V.~E.~Shemarulin, {\it Int. J. Mod. Phys. A} {\bf36}, 2150086 (2021), arxiv: 2108.04664.

\bibitem{bib13} V.~P.~Neznamov, {\it Int. J. Mod. Phys. A} {\bf36}, 2150173 (2021), arxiv: 2110.03530.

\bibitem{bib14} V.~P.~Neznamov, {\it Int. J. Mod. Phys. A} {\bf40}, 2550049 (2025).

\bibitem{bib15} V.~P.~Neznamov, {\it Int. J. Mod. Phys. A } {\bf40}, 2550074 (2025).

\bibitem{bib16} V.~P.~Neznamov, A.~J.~Silenko, {\it J. Math. Phys.} {\bf50}, 122302 (2009); arxiv: 0906.2069 (math-ph).

\bibitem{bib17} J.~D.~Bjorken, S.~D.~Drell, {\it Relativistic Quantum Mechanics,} (New York: McGraw-Hill ,1964).

\bibitem{bib18} W.~Greiner, J.Reinhardt, {\it Quantum Electrodynamics}(Berlin: Springer, 2002).

\bibitem{bib19} I.~Pomeranchuk, J.~Smorodinsky, {\it J. Phys. USSR} {\bf 9}, 97 (1945).

\bibitem{bib20} S.~S.~Gershtein, Y.~B.~Zel'dovich, {\it Sov. Phys. JETP} {\bf 30}, 358 (1970).

\bibitem{bib21} W.~Pieper, W.~Greiner, {\it Z. Phys.} {\bf 218}, 327 (1969).

\bibitem{bib22} Y.~B.~Zel'dovich, V.~S.~Popov, {\it Sov. Phys. Usp}. {\bf 14}, 673 (1972).

\bibitem{bib23} O'Connell R.~F. ''Rotation and spin in physics,'' in General Relativity and John Archibald Wheeler (Astrophysics and Space Science Library, Vol. 367, Eds I Ciufolini, R~A~Matzner) (Dordrecht: Springer, 2010) p. 325, DOI:10.1007/978-90-481-3735-0$_{-}$14.

\bibitem{bib24} O'Connell R.~F.~, {\it Mod. Phys. Lett. A} {\bf 26}, 469 (2011).

\bibitem{bib25} V.~P.~Neznamov, {\it FIZMAT}, V.2 94 (2024) (in Russian).

\bibitem{bib26} V.~P.~Neznamov, {\it Int. J. Mod. Phys. A} {\bf40}, 255010 (2025).

\bibitem{bib27} P.~A.~M.~Dirac, {\it Proc. Roy. Soc. A} {\bf 322}, 435 (1971).

\bibitem{bib28} P.~A.~M.~Dirac, {\it Proc. Roy. Soc. A} {\bf 328}, 1 (1972).

\end{thebibliography}
\end{document}